\documentclass[aps,prl,
superscriptaddress,
twocolumn
]{revtex4-1}

\usepackage{graphicx}
\usepackage{dcolumn}
\usepackage{bm}
\usepackage[mathlines]{lineno}
\usepackage[utf8]{inputenc}
\usepackage[T1]{fontenc}
\usepackage[english]{babel}
\usepackage{color}
\usepackage{hyperref}

\newcommand{\beq}{\begin{equation}} 
\newcommand{\eeq}{\end{equation}}
\newcommand{\bfig}{\begin{figure}}
\newcommand{\efig}{\end{figure}}
\newcommand{\igraph}{\includegraphics}

\graphicspath {{C:/Users/caspe/Dropbox/Dottorato/memorie/resume/proceeding/immagini/}}

\begin{document}


\title{Properties of ferromagnetic Josephson junctions for memory applications}

\author{R. Caruso}
\affiliation{Dipartimento di Fisica, Università di Napoli Federico II; Monte Sant'Angelo - via Cintia; I-80126 Napoli - Italy}
\affiliation{CNR-SPIN, Monte S. Angelo - Via Cintia, I-80126 Napoli - Italy}
\author{D. Massarotti}
\affiliation{Dipartimento di Ingegneria Elettrica e delle Tecnologie dell'Informazione, Università di Napoli Federico II, Via Claudio, I-80125 Napoli, Italy}
\author{A. Miano}
\affiliation{Dipartimento di Fisica, Università di Napoli Federico II; Monte Sant'Angelo - via Cintia; I-80126 Napoli - Italy}
\affiliation{CNR-SPIN, Monte S. Angelo - Via Cintia, I-80126 Napoli - Italy}
\author{V.V. Bol'ginov}
\affiliation{Institute of Solid State Physics (ISSP RAS), Chernogolovka, Moscow Region
142432 - Russia}
\affiliation{Skobeltsyn Institute of Nuclear Physics,  Moscow State University,  Moscow, 119991  Russia}
\author{A. Ben Hamida}
\affiliation{National University of Science and Technology MISIS, 4 Leninsky prosp., Moscow
119049 - Russia}
\affiliation{Leiden Institute of Physics, Leiden University, Niels Bohrweg 2, 2333 CA Leiden - The Netherlands}
\author{L.N. Karelina}
\affiliation{Institute of Solid State Physics (ISSP RAS), Chernogolovka, Moscow Region
142432 - Russia}
\author{G. Campagnano}
\affiliation{Dipartimento di Fisica, Università di Napoli Federico II; Monte Sant'Angelo - via Cintia; I-80126 Napoli - Italy}
\author{I.V. Vernik}
\affiliation{HYPRES, Inc. - 175 Clearbrook Road, Elmsford, NY 10523 - USA}
\author{F. Tafuri}
\affiliation{Dipartimento di Fisica, Università di Napoli Federico II; Monte Sant'Angelo - via Cintia; I-80126 Napoli - Italy}
\affiliation{CNR-SPIN, Monte S. Angelo - Via Cintia, I-80126 Napoli - Italy}
\author{V.V. Ryazanov}
\affiliation{Institute of Solid State Physics (ISSP RAS), Chernogolovka, Moscow Region
142432 - Russia}
\affiliation{Faculty of Physics, National Research University Higher School of Economics, Moscow - Russia
}
\author{O.A. Mukhanov}
\affiliation{HYPRES, Inc. - 175 Clearbrook Road, Elmsford, NY 10523 - USA}
\author{G.P. Pepe}
\affiliation{Dipartimento di Fisica, Università di Napoli Federico II; Monte Sant'Angelo - via Cintia; I-80126 Napoli - Italy}
\affiliation{CNR-SPIN, Monte S. Angelo - Via Cintia, I-80126 Napoli - Italy}



\begin{abstract}

In this work we give a characterization of the RF effect of memory switching on Nb-Al/AlO$_x$-(Nb)-Pd$_{0.99}$Fe$_{0.01}$-Nb Josephson junctions as a function of magnetic field pulse amplitude and duration, alongside with an electrodynamical characterization of such junctions, in comparison with standard Nb-Al/AlO$_x$-Nb tunnel junctions. The use of microwaves to tune the switching parameters of magnetic Josephson junctions is a step in the development of novel addressing schemes aimed at improving the performances of superconducting memories.

\end{abstract}


\maketitle


\section{Introduction}

The use of single flux quantum (SFQ) pulses instead of the voltage levels for superconducting digital logic was introduced in 1985 \cite{RSFQ1, RSFQ2}. In 1987-1989, the first Rapid Single Flux Quantum (RSFQ) integrated circuits were developed\cite{RSFQ3, RSFQ4}. Although RSFQ logic could operate at an extremely high clock speed with good operating margins, its energy efficiency was limited by static power dissipation due to the use of bias resistors. As a result, RSFQ circuits could hardly be scalable for large scale computing applications or very low power and ultra-low temperature applications such as required for the readout/control circuits in quantum computing or cryogenic detector arrays. In order to reduce the static power dissipation, several different approaches have been developed \cite{LR-RSFQ1, LVRSFQ1, LVRSFQ2, RQL, ERSFQ2, ERSFQ1, eSFQ, ULPSFQ}. In particular, zero-static power dissipation ERSFQ logic allows the realization of circuits of increased complexity \cite{ERSFQ3} and can address the energy dissipation challenges that is now facing traditional large scale computers based on conventional complementary metal-oxide semiconductor (CMOS) circuits. 
However, the progress in superconducting random access memories (RAM) indispensable for large scale computing applications was significantly slower. This encouraged an active research in new memory devices for the implementation of a cryogenic, dense, energy-efficient RAM compatible to energy-efficient SFQ logics \cite{Jmagn, memIgor, memorieJAPnoi, memorie, sfsmem1, sfsmem6, spinvalve, sfsmem7, memGoldobin}. A successful realization of such a RAM would require not just small and low-power memory elements, but also energy- and area-efficient addressing approaches in RAM arrays. Otherwise, the RAM power and density will be determined by the read/write addressing circuits.

Here we focus on the effect of RF pulses on switching processes of low dissipation magnetic Josephson junctions (MJJs), composed by niobium electrodes and a multilayered barrier with an Al/AlO$_x$ insulating layer, a thin niobium interlayer, and a weak ferromagnetic Pd$_{0.99}$Fe$_{0.01}$ layer. 
It has been demonstrated that in such junctions it is possible to switch between two states with significantly different critical current values using magnetic field pulses. In Fig. \ref{fig:ich} (a) and (b) we show the two different I-V curves obtained for $H = 0.6 mT$, corresponding to the blue and green $I_C (H)$ curves respectively in Fig. \ref{fig:ich} (c).
If the initial state is represented in Fig. \ref{fig:ich} (a), the memory element can be switched into the state in Fig. \ref{fig:ich} (b) using a positive field pulse. On the rising edge of the pulse, the critical current moves along the blue curve in Fig. \ref{fig:ich} (c). On the falling edge of the pulse, the critical current follows the green curve, and after the end of the pulse, the junction ends up in the state in Fig. \ref{fig:ich} (a). The hysteresis of the $I_C (H)$ curves depending on the magnetic field ramp is due to the ferromagnetic barrier\cite{Jmagn,memIgor}, and it can be used to calculate the $M(H)$ curve shown in Fig.\ref{fig:ich} (d).
In order to read the logic state, we use a bias current between the two critical currents: when the junction is in the '0' state (high critical current state, Fig. \ref{fig:ich} a), there is no voltage across the junction. On the other hand, when the junction is in the '1' state (low critical current state, Fig. \ref{fig:ich} b), a voltage appears across the junction.

We have demonstrated that the combined application of an RF signal together with magnetic field pulses enhances the separation between critical current levels\cite{memorieJAPnoi}. This effect is related to the damping of the coercive field of the ferromagnetic barrier caused by the excitation of the magnetic moments due to microwave signal. In this way it is possible to select the amplitude of the writing field pulse in such a way that only the section of the memory array subject to RF field can change its digital state. 
This is the first step on the path of developing alternative schemes to manipulate the memory states of cryogenic RAMs, which are crucial in order to achieve higher density and efficiency.

In this work we give an electrodynamical characterization of MJJ junctions and study the effect of RF fields on switching processes as a function of different parameters such as magnetic field pulse amplitude and pulse duration.

\bfig
\centering
\igraph[width=\columnwidth]{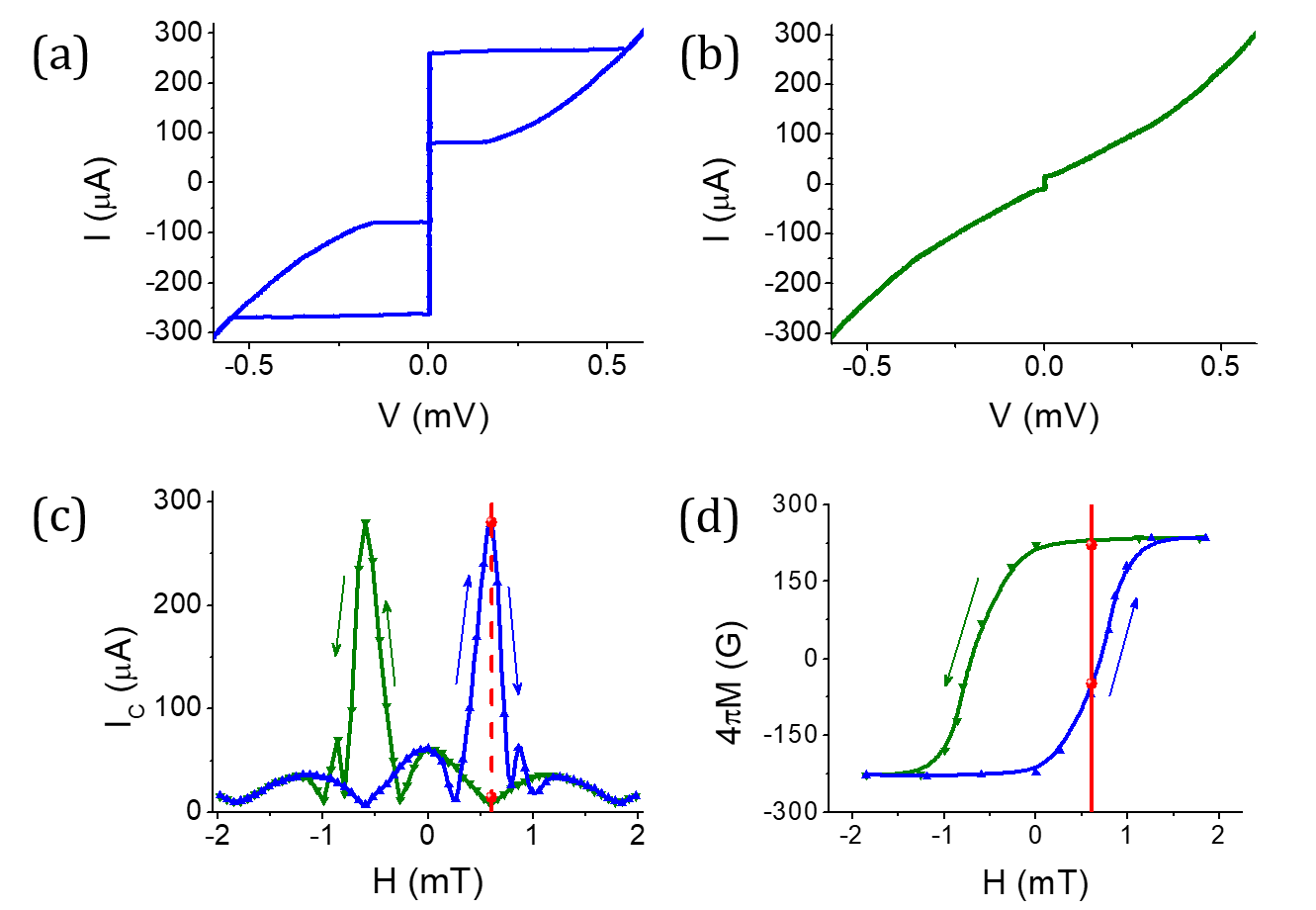}
\caption{(a) and (b) I-V curves measured in the high critical current state and in the low critical current state indicated by red dots in panel c. (c) $I_C (H)$ curves at 4 K for a junction with PdFe thickness $\approx$ 18 nm, measured ramping the magnetic field from negative to positive values (blue curve) and vice-versa (green curve). The red vertical line corresponds to the optimal working point. (d) $M(H)$ curve obtained by $I_C (H)$. All measurements have been performed at 4.2 K.}
\label{fig:ich}
\efig

\section{Methods}

The samples analyzed in this work have been realized within a collaboration between Hypres Inc. and ISSP\cite{memorie}. The bottom Nb-Al/AlO$_{x}$-Nb trilayer has been fabricated by Hypres using standard process to attain 4.5 $kA/cm^2$ critical current density\cite{hypresSTD2005, hypresSTD2007}, while the Pd$_{0.99}$Fe$_{0.01}$-Nb bilayer has been fabricated by ISSP. More details on the fabrication process can be found elsewhere\cite{memorie}.

\bfig
\begin{center}
\igraph[width=\columnwidth]{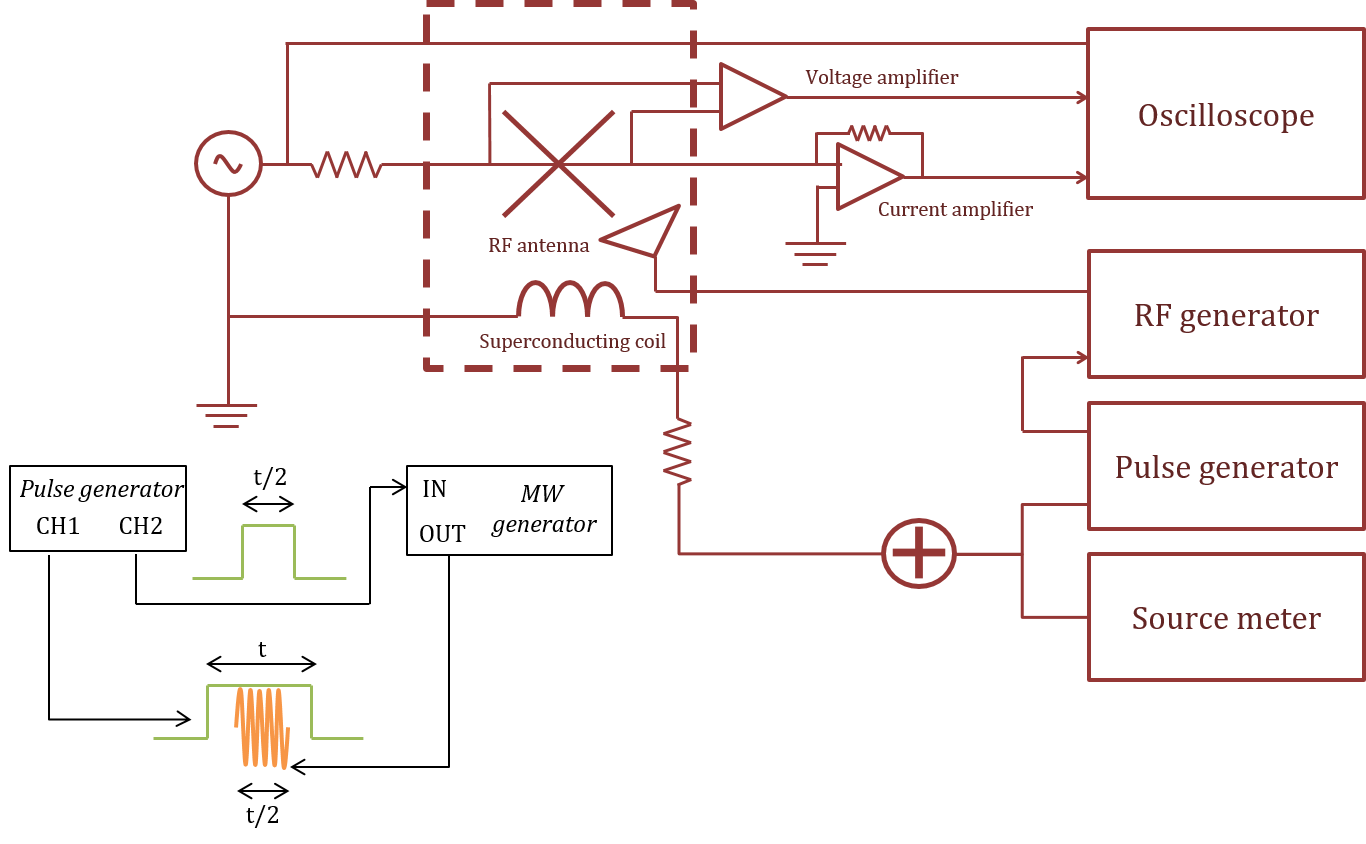}
\caption{Scheme of the measurement setup and driving pulse. The dashed box indicates the low temperature portion of the experiment. Bottom left: scheme of the addressing pulse.}
\label{fig:pulse}
\end{center}
\efig

The measurements have been performed using a Heliox-VL evaporation cryostat equipped with an RF antenna close to the sample stage and a NbTi superconducting coil used to apply magnetic field in the plane of the junction. The sample stage is thermally anchored to the $^3$He pot, while the superconducting coil is thermally anchored to the 1K-pot, in order to avoid sample heating. RC filters and copper powder filters with cutoff frequencies of $1 MHz$ and $1 GHz$ respectively are anchored to different thermal stages, in order to ensure optimal noise reduction\cite{luigiAPL2011, luigiPRB2011}. 
The junction is biased with a low frequency current ramp (approximately $11 Hz$) using a waveform generator in series with a shunt resistance, and the voltage across the junction is measured using a battery powered amplifier. 
 
In Fig. \ref{fig:pulse} we show a schematic representation of the measurement setup, while in the inset we focus on the memory driving signal composed by the magnetic field pulse and the RF train. The working point is set using a dc voltage signal, which is combined with a voltage pulse generated by one of the channels of the pulse generator using an adder. The output signal of the adder is then sent to a shunt resistance, so that the coil is current biased. The second channel of the pulse generator is used to drive the RF train (see inset in Fig.\ref{fig:pulse}), the length is set by the length of the driving pulse, while frequency and power level of the microwaves are controlled independently using the RF generator.  
We use a field pulse length of $500 ms$, with a rise and fall time of $1 ms$. The length required to induce the memory switch with our setup is quite  large because of the large characteristic time of the superconducting coil, which is approximately $10 ms$. The microwave train is centered around the center of the magnetic field pulse.
Given the high $I_C R_N$ product, MJJs are sought to operate at much higher speed, with much shorter pulses, when properly coupled with address and read-out circuitry.

In order to measure the current level separation we define $\Delta I$ as
\beq
\Delta I = (I_{C}^{high}-I_{C}^{low})/I_{C}^{high}
\label{eqn:deltaI}
\eeq

where $I_{C}^{high}$ is the critical current corresponding to '0' logic state and $I_{C}^{low}$ is the critical current corresponding to '1' logic state. Both $I_{C}^{high}$ and $I_{C}^{low}$ are obtained from an average of at least ten I-V curves obtained in the same conditions. To measure the enhancement in current level separation we define $N$ as

\beq
N=(\Delta I_{MW}-\Delta I_{noMW})
\label{eqn:G}
\eeq

where $\Delta I_{MW}$ is the current level separation in presence of microwaves and $\Delta I_{noMW}$ is in absence of microwaves.

All measurements presented in the following have been performed with a constant magnetic field bias of $1.2 G$, in order to set the optimal working point\cite{memorieJAPnoi}. This can be determined from the $I_C (H)$ curves measured ramping magnetic field from negative to positive values and vice-versa. 
The optimal working point coincides with the magnetic field value corresponding to the maximum separation between $I_C (H)$ curves (see Fig.\ref{fig:ich} c).

\section{Results}

\subsection{Comparison between MJJ and SIS}

\bfig
\centering
\igraph[width=\columnwidth]{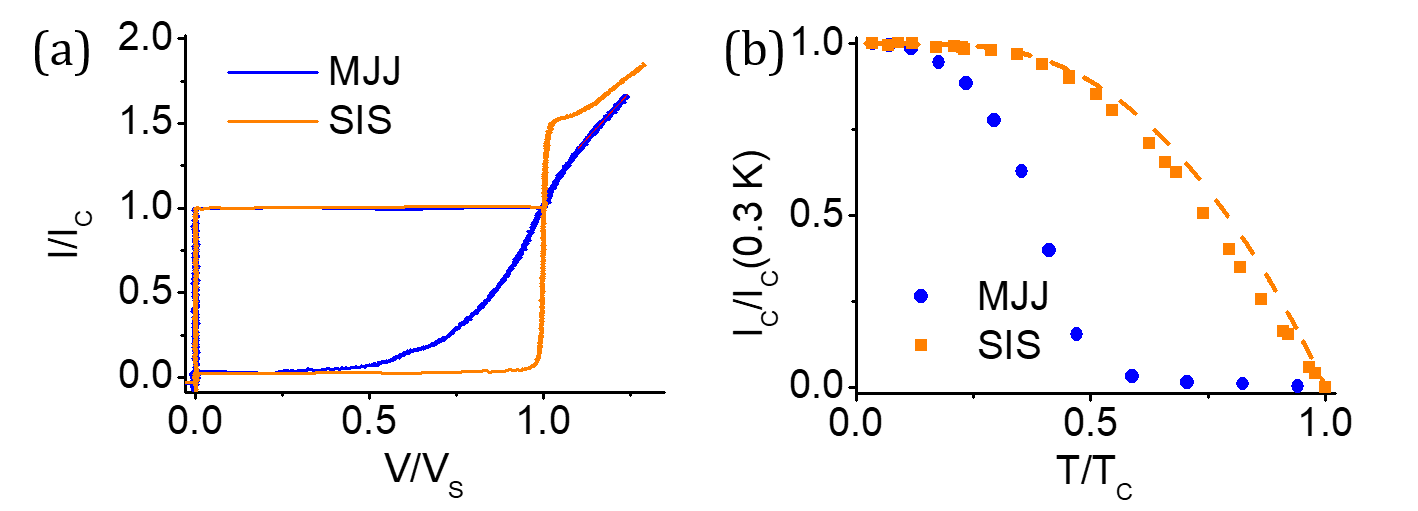}
\caption{I-V curves at $0.3 K$ (a) and $I_C(T)$ (b) for a magnetic Josephson junction (blue) with a 18 nm thick PdFe layer and a standard Nb-AlAlO$_x$-Nb tunnel junction (orange). Orange dashed line is an Ambegaokar-Baratoff fit of the Nb-AlAlO$_x$-Nb junction. The estimated critical temperature is $8.6\pm0.2 K$.}
\label{fig:comp}
\efig

A comparison can be done between MJJs with standard SIS junctions with Nb electrodes and Al/AlO$_x$ barrier fabricated using standard Hypres technology\cite{hypresSTD2005}.
As reported in previous works\cite{memorie}, MJJ are compatible in speed with existing RSFQ circuitry thanks to their high $I_C R_N$ product, which is $\approx 700\mu V$\cite{memorie} for MJJ and $\approx 1.5 mV$ for SIS.  

When a Josephson junction is embedded in a circuit, it can be modeled as an ideal Josephson junction in parallel with a resistance $R$ and a capacitance $C$, according to the Resistively and Capacitively Shunted Junction model (RCSJ)\cite{barone, likharevBook}. The balance equation for this circuit can be rearranged using the two Josephson equations into a motion equation for a phase particle moving in a tilted washboard potential and subject to a viscous force:
\beq
\frac{d^2\varphi}{d\tau^2}+\frac{1}{Q}\frac{d\varphi}{d\tau}=-\sin\varphi+\frac{I(\varphi)}{I_{C}}
\label{eqn:RCSJ}
\eeq
where $\tau$ is the normalized time with respect to plasma frequency $\omega_P = \sqrt{\frac{2eI_C}{\hbar C}}$, $\varphi$ is the phase difference between the two electrodes, $Q$ is the junction quality factor (also called damping factor) and $I_{C}$ is the critical current. The damping factor is defined as
\beq
Q=\omega_P R C
\label{eqn:Q}
\eeq
where $\omega_P=\sqrt{\frac{2eI_C}{\hbar C}}$ is the plasma frequency. If $Q\gg 1$,
then a finite capacitance is associated with the dielectric barrier. As reported elsewhere\cite{zappe}, it is possible to estimate $Q$ from the hysteresis of the I-V curves. 
For tunnel junctions with small, flat subgap currents in a large voltage interval, as in the case of SIS junction in Fig. \ref{fig:comp} a, this method fails to give a reliable estimation of the damping factor $Q$. In such conditions, the method proposed is very sensitive to small variations of the hysteresis parameter, and a small uncertainty on the retrapping current results in a large uncertainty on the quality factor.

The resistance $R$ appearing in Eq. \ref{eqn:Q} is voltage-dependent, and takes into account the non-linear resistance in the I-V curve and the external resistors in parallel with the junction itself\cite{zappe}. In the zero voltage state, the junction is equivalent to a phase particle oscillating in one of the minima of the washboard potential with a frequency given by $\omega_P$. The plasma frequency is typically around a few $GHz$. At such frequencies, the dissipation is dominated by the external impedance, which is typically much smaller than the junction resistance. The total impedance due to the external circuitry is usually around 100$\Omega$ \cite{dev1}.
Moreover, it is well known that in standard tunnel Nb-Al/AlO$_x$-Nb junctions the specific capacitance depends on the critical current density through the empirical relation\cite{capacitance}
\beq
\frac{1}{C_s}=0.2-0.043 \log_{10} J_c
\label{eqn:Cs}
\eeq
where $C_s$ is expressed in $\mu F/cm^2$ and $J_c$ in $A/cm^2$. Substituting the nominal value\cite{hypresSTD2005} $J_c = 4.5 kA/cm^2$ we obtain $C\approx 6 pF$. Using $R\approx 100\Omega$, we obtain $Q\approx 230$ for the SIS junction.

The presence of a metallic PdFe barrier in the MJJ leads to a different behavior of the subgap current: at 1 mV, we observe $I = 8\mu A$ for the SIS junction and $I=39 \mu A$ for the MJJ. These values, together with the overall trend of the subgap current, indicate a higher dissipation in MJJ junctions when compared with standard SIS junctions, still preserving an underdamped behavior with a large quality factor.  
The larger dissipation and the subsequent different trend of the subgap current in MJJ allows us to use the hysteresis of the I-V curve to estimate the quality factor of the junction. This estimation gives $Q\approx 40$. 
The method we used to estimate the capacitance in SIS using Eq. \ref{eqn:Cs} cannot be used for MJJ, due to the presence of the ferromagnetic layer and of the additional superconducting interlayer, which change the overall capacitance of the junction. We can estimate the capacitance from the damping factor $Q$, taking into account that, due to the multilayered structure of the barrier with a metallic layer, the resistance appearing in Eq. \ref{eqn:Q} is no longer the lead impedance but the junction normal state resistance, $R_N\sim 5\Omega\ll 100\Omega$\cite{golubovSIFS}.
The estimated capacitance for the MJJ is then $C=9\pm 3 pF$. The different capacitance, compared to SIS junction capacitance, is due to the different layers involved in the barrier structure and to the larger area of the MJJ junction, together with the larger separation between the two superconducting electrodes.

In Fig. \ref{fig:comp} b we show $I_C (T)$ curves for the same two junctions. The critical current temperature behavior of the SIS junction resembles the well known Ambegaokar-Baratoff formula\cite{AB, ABerr}, while the experimental data for MJJ are in agreement with the model for SIsFS junctions in literature\cite{baku}, with a tunnel-like behavior at low temperature and a pronounced proximity effect tail at higher temperatures.

\subsection{Shapiro steps}

\bfig
\centering
\igraph[width=0.7\columnwidth]{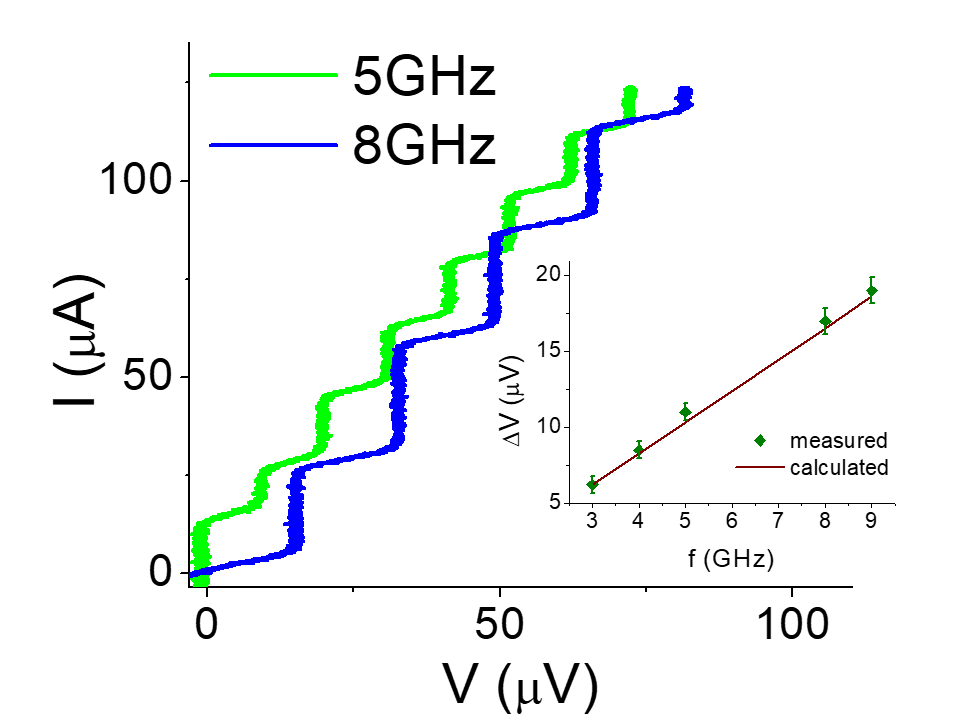}
\caption{I-V curves showing Shapiro steps for different microwave frequencies. Inset: width of the Shapiro voltage steps as a function of frequency. Green diamonds are measured values, dark red line represents the expected values calculated using \ref{eqn:shapiro}}
\label{fig:shapiro}
\efig

The appearance of current steps in the I-V characteristic in presence of an external RF field is a well known consequence of the second Josephson equation\cite{shapiro}. The voltage steps appear at \cite{barone}
\beq
V_n=\frac{n h}{2 e}\nu
\label{eqn:shapiro1}
\eeq
where $n$ is an integer, $e$ is the electron charge and $h$ is the Planck constant. The ratio between frequency and voltage is then given by\cite{barone}
\beq
\frac{\nu}{V}=483 \mbox{ } MHz/\mu V
\label{eqn:shapiro}
\eeq

We measured such steps for different frequencies of applied microwaves, for a MJJ with $18 nm$ thick PdFe at $4.2 K$. In Fig. \ref{fig:shapiro} we show I-V curves in presence of an external microwave field at $5 GHz$ (green line) and $8 GHz$ (blue line), and in the inset we show the measured voltage step (green diamonds) and its expected value (dark red line) as a function of microwave frequency.

The appearance of Shapiro steps in the I-V characteristic indicates a good coupling between the junction and the external RF field, and thus allows us to identify the optimal working frequency to observe current level separation enhancement in switching processes. 

All the measurements presented here have been performed at $3.88 GHz$, close to the frequency where Shapiro steps have been observed for the sample with $14 nm$ thick PdFe barrier.
The working frequency we identified is also close to ferromagnetic resonance (FMR) frequency measured for thicker PdFe thin films \cite{resonances}. 

The use of a RF frequency close to a frequency at which Shapiro steps are observed is a way to ensure a certain degree of interaction between the microwaves and the sample. In principle, Shapiro steps should be observed in the whole microwave frequency range. However, in our setup this no longer holds true, as the coupling between the external field and the sample depends on several factors such as the distance and relative position of the emitting antenna and the sample. Only certain frequency ranges allow to observe Shapiro steps, we use such frequencies in our experiment in order to be sure that there is an actual interaction between the RFs and the sample.

\subsection{Pulse amplitude and length dependence}

\bfig
\centering
\igraph[width=\columnwidth]{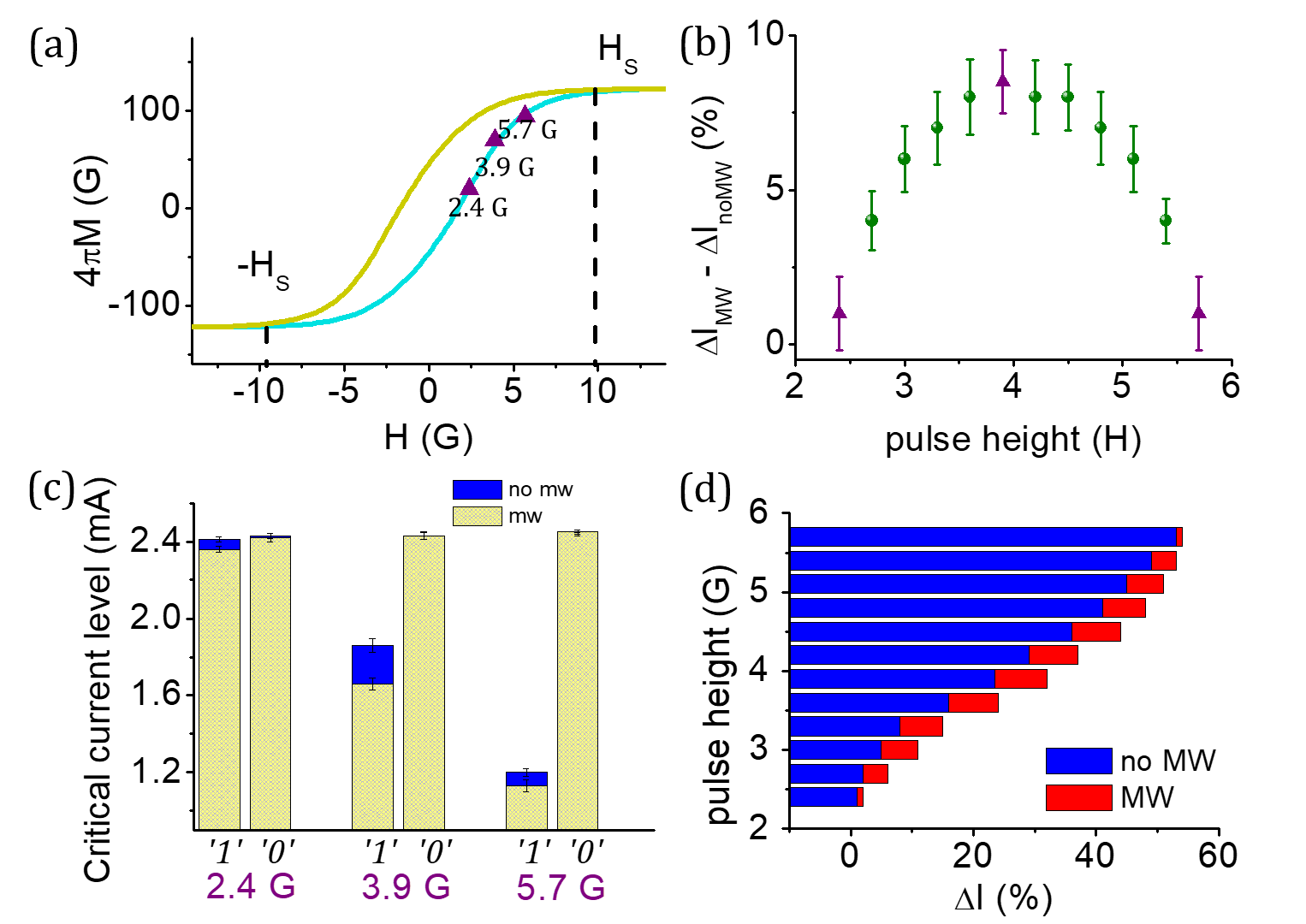}
\caption{(a) Magnetization curve for a sample with PdFe thickness $\approx 14 nm$. Purple triangles correspond to the magnetic field pulse amplitudes where the level separation enhancement ($G$) is minimum ($2.4 G$ and $5.7 G$) and maximum ($3.9 G$) (b) Level separation enhancement for different values of the magnetic field pulse amplitude. Purple triangles correspond to field pulse amplitudes indicated by purple triangles in panel a. (c) Critical current levels for the three values of magnetic pulse amplitude highlighted in panels a and b, in presence and in absence of external RF field. (d) Current level separation ($\Delta I$) for different values of the magnetic field pulse amplitude in absence and in presence of microwaves. }
\label{fig:G}
\efig

We measured $N$ for different values of the magnetic field pulse amplitude, fixing frequency, power level and pulse length. We used $500 ms$ long field pulses combined with $250 ms$ long RF trains, centered around the center of the field pulse. The frequency of the RF train is $3.88 GHz$, while the power level is $4.9 dBm$. We observed that $N$ is almost zero for small values of the magnetic field, around $2.4 G$, then it increases up to its maximum value at $3.9 G$ and finally goes to zero at approximately $5.4 G$, as shown in Fig. \ref{fig:G} (d). It has been shown \cite{memorieJAPnoi} that this behavior can be explained considering the magnetization curve of the ferromagnetic barrier, obtained using Josephson magnetometry \cite{Jmagn} in Fig. \ref{fig:G} (a). For small field pulse amplitudes $M(H)$ curve is almost linear, and so there is no difference between high and low critical current levels within the error bars, and the application of an external RF field does not change it significantly. At large pulse amplitudes, the ferromagnet is close to its saturation, and so the influence of external RF fields is negligible. Intermediate fields correspond to an intermediate region of $M(H)$ curve, and so here the effect of microwaves is most significant. This results in an optimal working point to observe the microwave effect on MJJs, depending on the $M(H)$ curve of the sample. For the measurement presented in Fig. \ref{fig:G} and \ref{fig:energy}, the optimal field amplitude is $3.9 G$.

In Fig. \ref{fig:G} (c) we show critical current levels for the two logical states of the memory at different values of the field pulse amplitude. The blue bars correspond to the current levels in absence of microwaves, while the yellow ones are collected when RF trains are applied. The high critical current levels are less affected by the microwaves because they correspond to the region of the $M(H)$ curve in the vicinity of the coercive field $H_C$, where $M\approx 0$. In such region, the small changes in $M(H)$ induced by the external RF field do not affect the observed current. On the other hand, the low critical current levels correspond to large $M$, and so the changes induced by microwaves become more evident.

\bfig
\centering
\igraph[width=\columnwidth]{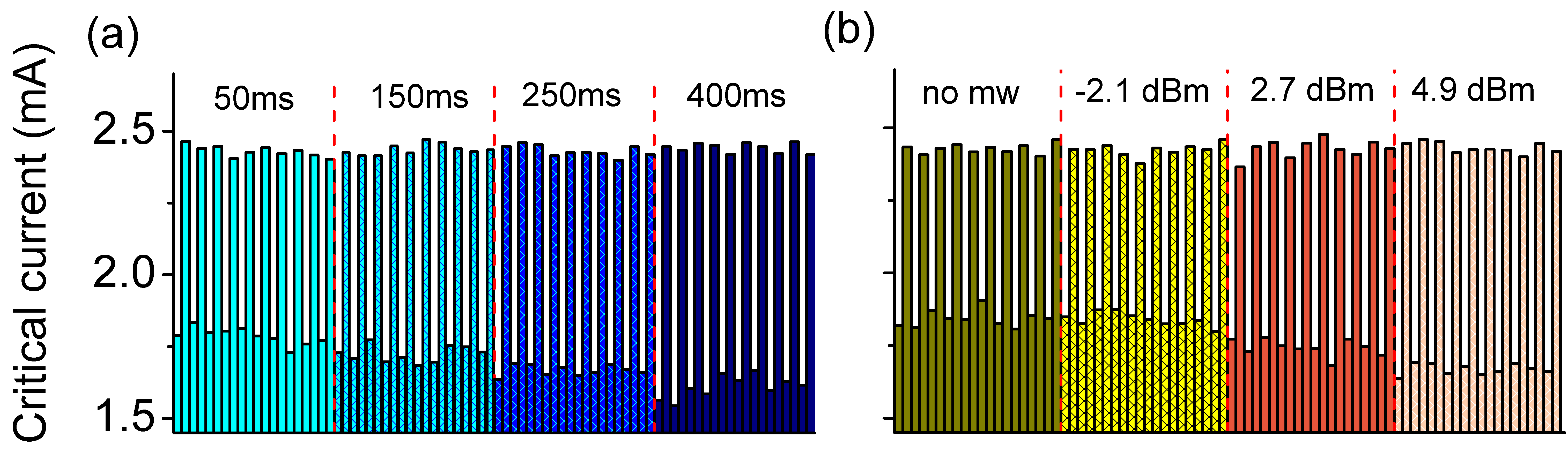}
\caption{(a) Critical current levels for RF trains with different duration at $4.9 dBm$. (b) Critical current levels for RF trains with different power levels and $250 ms$ duration.}
\label{fig:energy}
\efig

The observed current level separation increases when increasing the time duration or the power level of applied RF train. In Fig.\ref{fig:energy} (a) we show the the measured current levels with fixed microwave power level ($4.9 dBm$) and different time duration: $50 ms$, $150 ms$ and $400 ms$. 
As reported in section II, the length of the field pulses is set by our measurement setup rather than the intrinsic switching time of the MJJ. In Fig. \ref{fig:energy} (b) we report the current levels obtained for different power level values with a time duration of the RF pulse set to $250 ms$.
This supports our previous conclusions\cite{memorieJAPnoi} on the dependence of the RF effect from the energy transferred to the junction. 

In particular, we identified the optimal working point to be used to test alternative addressing schemes for magnetic memories based on Josephson junctions using more appropriate circuits, such as coplanar waveguides.

\section{Conclusion}

We characterized for the first time magnetic Josephson junctions down to $0.3 K$, and determined the electrodynamical parameters of the junction, which is an important step towards the integration of these devices in complex circuits. We report measurements of ac Josephson effect in presence of microwaves that allow us to identify the optimal microwave frequencies to observe the current level separation enhancement in our setup. We also present measurements supporting our previous conclusions on the effect of RF fields on memory switching processes. 
We have also shown that the perturbative effect of microwaves becomes important for large $M$, and thus it is observed in particular on the low critical current states of MJJs. 
These results play an important role for future implementation of RF-based addressing schemes for MJJs and potentially lead to higher hardware-, area-, and energy-efficient cryogenic memory solutions. 

\section{Acknowledgments}
This work has been partially financed by \emph{B5 2F17 001400005} grant within the framework of SEED 2017 of CNR-SPIN. V.V. Bolginov and L.N. Karelina thank Russian Foundation for Basic Research grant \emph{17-02-01270}, A. Ben Hamida acknowledges the Ministry of Education and Science of the Russian Federation in the framework of Increase Competitiveness Program of NUST "MISiS" (research project N.\emph{K4-2014-080}), V.V. Ryazanov thanks joint Russian-Greece grant \emph{2017-14-588-0007-011}. The authors would also like to thank NANOCOHYBRI project (Cost Action CA 16218).

\end{document}